\documentclass[	aps,
		prl,
		twocolumn,
		a4paper,
		superscriptaddress,
		10pt
]{revtex4-1}

\usepackage{graphicx}
\usepackage{amsmath}
\usepackage{amssymb}
\usepackage{xcolor}
\usepackage[colorlinks=true,urlcolor=blue,citecolor=blue,linkcolor=magenta]{hyperref}
\usepackage{ulem}
\usepackage{braket}

\renewcommand{\rm}[1]{\mathrm{#1}}

\begin{document}
\title{Conditional Hybrid Nonclassicality}

\author{E. Agudelo}\email{elizabeth.ospina@uni-rostock.de}
\affiliation{AG Theoretische Quantenoptik, Institut f\"ur Physik, Universit\"at Rostock, D-18051 Rostock, Germany}

\author{J. Sperling}
\affiliation{Clarendon Laboratory, University of Oxford, Parks Road, Oxford OX1 3PU, United Kingdom}

\author{L. S. Costanzo}
\affiliation{Istituto Nazionale di Ottica (INO-CNR), Largo Enrico Fermi 6, 50125 Florence, Italy}
\affiliation{LENS and Deparment of Physics, University of Firenze, 50019 Sesto Fiorentino, Florence, Italy}

\author{M. Bellini}
\affiliation{Istituto Nazionale di Ottica (INO-CNR), Largo Enrico Fermi 6, 50125 Florence, Italy}
\affiliation{LENS and Deparment of Physics, University of Firenze, 50019 Sesto Fiorentino, Florence, Italy}

\author{A. Zavatta}
\affiliation{Istituto Nazionale di Ottica (INO-CNR), Largo Enrico Fermi 6, 50125 Florence, Italy}
\affiliation{LENS and Deparment of Physics, University of Firenze, 50019 Sesto Fiorentino, Florence, Italy}

\author{W. Vogel}
\affiliation{AG Theoretische Quantenoptik, Institut f\"ur Physik, Universit\"at Rostock, D-18051 Rostock, Germany}

\begin{abstract}
	We derive and implement a general method to characterize the nonclassicality in compound discrete- and continuous-variable systems. 
	For this purpose, we introduce the operational notion of conditional hybrid nonclassicality which relates to the ability to produce a nonclassical continuous-variable state by projecting onto a general superposition of discrete-variable subsystem. 
	We discuss the importance of this form of quantumness in connection with interfaces for quantum communication. 
	To verify the conditional hybrid nonclassicality, a matrix version of a nonclassicality quasiprobability is derived and its sampling approach is formulated. 
	We experimentally generate an entangled, hybrid Schrödinger cat state, using a coherent photon-addition process acting on two temporal modes, and we directly sample its nonclassicality quasiprobability matrix. 
	The introduced conditional quantum effects are certified with high statistical significance.
\end{abstract}

\date{\today}
\maketitle

\paragraph*{Introduction.---}
	The investigation of signatures of nonclassicality is of crucial importance for the understanding, engineering, and control of quantum systems.
	The knowledge about various forms of quantumness plays a central role in modern research, ranging from fundamental tests of quantum physics \cite{reid09, hensen15} to applications close to commercial quantum information processing \cite{aaa,bbb,ccc}.
	Especially for secure communication, quantum correlations of light have been vastly exploited \cite{scarani09}.
	To apply quantum enhanced communication protocols, it is indispensable to characterize the correlations between different systems.
	In particular, the interface between continuous-variable (CV) and discrete-variable (DV) systems has to be understood for implementing quantum communication on a highly variable basis and for employing the benefits of both kinds of systems \cite{gisin07}.

	In CV quantum optics, the standard concept of nonclassicality is based on the impossibility of describing field correlations in terms of classical electrodynamics.
	This notion defines nonclassical light via a Glauber-Sudarshan $P$ representation \cite{glauber63, sudarshan63,titulaer65} that does not resemble a classical probability distribution.
	Thus, quasiprobability representations are a direct and intuitive way to discern classical from quantum field theories.
	Moreover, the negativities of these quasiprobabilities have been closely related to other features, like contextuality \cite{spekkens08,ferrie09} and symmetries of the quantum state \cite{zhu16}.

	However, the $P$ distribution can be strongly singular for many states \cite{sperling16}.
	Thus, different strategies have been investigated in order to regularize the $P$ distribution.
	For instance, $s$-parametrized quasiprobabilities \cite{cahill69a, cahill69b} have been introduced, which include the Wigner function for $s=0$.
	But the $s$~parameter not only regularizes the singularities, it also limits the sensitivity to verify quantumness.
	To overcome this deficiency, non-Gaussian nonclassicality filters have been applied to uncover all forms of single- and multimode nonclassicality via regular nonclassicality quasiprobabilities \cite{kiesel10,agudelo13}.

	The DV regime of quantum optics can be, for example, related to the particle picture of quantized fields using the Fock representation of states.
	This expansion in terms of photon number states also allows for a complete characterization of light fields \cite{vogel_book}.
	Other realizations of such so-called photonic qudits are based on the angular momentum of light \cite{malik14,bent15}.
	The advantage of this representation is clearly its direct connection to quantum information processing, which is formulated in qudits as the basic carriers of information.

	Traditionally, the CV and DV representation of light have been individually exploited, but the connection between these two complementary regimes has not been extensively studied.
	One of the few relations between CV and DV systems that has been established is based on the observation that qubits can be constructed out of CV states \cite{neergaard10,coelho16}.
	Another early attempt to consider a hybrid system was elaborated to measure nonclassicality between a vibrational mode and the electronic states of a trapped two-level atom based on a Wigner matrix \cite{wallentowitz97}.
	More recently, the experimental generation of hybrid entanglement was reported \cite{jeong14, morin14} and its entanglement had been investigated \cite{kreis12,constanzo15}.
	Even though quantum-correlated hybrid systems are of high interest for quantum information processing \cite{vanloock11,andersen15}, a universal way to access their nonclassicality is missing yet.

	In this Letter, we introduce and implement a general method for describing and identifying nonclassicality in hybrid, i.e., correlated CV and DV systems of light.
	The operational meaning of the resulting notion of conditional hybrid nonclassicality (CHN) is motivated in terms of resources for quantum communications.
	In contrast to previous approaches, formulated in terms of joint correlations, our technique is based on the conditional nonclassicality which directly relates to remote state preparation and manipulation protocols.
	Our concept of CHN is shown to be experimentally accessible via a nonclassicality quasiprobability (NQP) matrix, which completely describes the hybrid system.
	The sampling theory for this NQP matrix is derived, which applies to imperfect data sets and, thus, goes beyond previously known state-reconstruction methods.
	To demonstrate the experimental application, we realize the prominent example of an entangled, hybrid Schr\"odinger cat state.
	From our data, we reconstruct the NQP matrix and certify its CHN with high statistical significance.

\paragraph*{Conditional hybrid nonclassicality.---}
	We firstly consider an application to motivate our general concept of CHN.
	Suppose we have a Schr\"odinger cat state,
	\begin{align}\label{eq:new_CatState}
		|\Psi_\mathrm{cat}\rangle=2^{-1/2}(|\beta\rangle\otimes|0\rangle+|{-}\beta\rangle\otimes|1\rangle).
	\end{align}
	The first subsystem is given in a CV description of coherent states.
	For the second subsystem, we have a DV expansion in terms of Fock states.
	The implementation of $|\Psi_\mathrm{cat}\rangle$ was established through tailoring the correlation between $|\pm\beta\rangle$ and $|0\rangle,|1\rangle$~\cite{wineland,haroche,jeong14,kwon13,lee13}.

	For establishing a DV-CV communication node, we aim at transferring the information of a qubit $|q\rangle=q_0|0\rangle+q_1|1\rangle$ (a third subsystem) into a CV encoding via the state \eqref{eq:new_CatState}.
	This can be achieved by performing a joint projection of the composed state $|\Psi_\mathrm{cat}\rangle\otimes|q\rangle$ onto the Bell state $|\psi\rangle=2^{-1/2}(|0\rangle\otimes|0\rangle+|1\rangle\otimes|1\rangle)$ in the second and third mode as it is done in quantum teleportation protocols \cite{BB93}.
	Hence, we get analogously
	\begin{align}\label{eq:new_Teleportation}
		|\Psi_q\rangle=\mathcal N(q_0|\beta\rangle+q_1|{-}\beta\rangle),
	\end{align}
	where $\mathcal N$ is the normalization constant; see Ref. \cite{ulanov17} for a different implementation.
	The state \eqref{eq:new_Teleportation} now carries the information of the qubit $|q\rangle$.
	If the qubit represents a classical truth value, $|q\rangle\in\{|0\rangle,|1\rangle\}$, we also get a classical coherent state $|{\pm}\beta\rangle$ in the resulting CV state \eqref{eq:new_Teleportation}.
	Yet, any superposition state $|q\rangle$ will also result in a nonclassical superposition state $|\Psi_q\rangle$.
	In other words, the hybrid state \eqref{eq:new_CatState} has the potential to yield a nonclassical CV state through DV projections.

	Let us abstract the above observation.
	Suppose $\hat\rho$ is a CV-DV-hybrid state and $\hat \Pi$ is a non-negative projection operator in the $d$-dimensional DV subsystem.
	Without loss of generality we can restrict ourselves to rank-one operators $\hat\Pi=|\psi\rangle\langle\psi|$, with $|\psi\rangle=\sum_{m=0}^{d-1} \psi_m |m\rangle$ \cite{comment1}, because any other $\hat \Pi$ can be considered as a positive linear combination of rank-one operators.
	Now, the conditional CV state is defined as
	\begin{align}\label{eq:new_ConditionalState}
		\hat\rho|_{\hat\Pi}=\mathcal N\mathrm{tr}_\mathrm{DV}(\hat\rho[\hat 1\otimes\hat \Pi])=\int d^2\alpha\, P(\alpha|\hat\Pi){|\alpha\rangle\langle\alpha|},
	\end{align}
	where $\mathcal N$ is a normalization constant, $\mathrm{tr}_\mathrm{DV}$ is the partial trace over the DV subspace, and $P(\alpha|\hat\Pi)$ is the conditional Glauber-Sudarshan distribution.
	We can now extend the concept of single-mode CV nonclassicality \cite{titulaer65} to define the operational notion of CHN:
	{\it The CV-DV hybrid state $\hat\rho$ shows CHN if there exists a DV projection $\hat\Pi$ such that $P(\alpha|\hat\Pi)$ is not a classical probability distribution.}

	In other words, CHN is the ability of a system to yield a nonclassical CV state for at least one measurement $\hat\Pi$ in the DV subsystem, $P(\alpha|\hat\Pi)\ngeqq0$.
	In particular, this definition applies to the Schr\"odinger cat state \eqref{eq:new_CatState} \cite{comment2}.
	Moreover, it extends the initial idea to mixed states and it generalizes DV systems beyond qubits.
	For example, a CV-qudit hybrid state $|\Psi\rangle=\sum_{m=0}^{d-1} c_m |\beta_m\rangle\otimes|m\rangle$ clearly exhibits CHN for nontrivial coefficients $c_m$.
	The entanglement and mixtures of such states have been extensively studied in Ref. \cite{kreis12}.
	In fact, any entangled hybrid state shows CHN, but CHN exists beyond entanglement.
	This results from the fact that CHN can be interpreted in terms of the heralded generation of nonclassical states which can be achieved with nonentangled states \cite{supplement}.
	Also note that another form of conditional nonclassicality has been recently studied in the context of photon statistics \cite{sperling16a}.

\paragraph*{Nonclassicality quasiprobability matrix.---}
	Verifying CHN requires one to explore all possible projections $\hat\Pi=|\psi\rangle\langle\psi|$.
	For at least one of them, the nonclassicality has to be verified from the conditional and possibly highly singular Glauber-Sudarshan distribution $P(\alpha|\hat\Pi)$.
	To overcome these difficulties, let us formulate a directly accessible and equivalent method.

	Using the concept of the $P$ representation, one can expand any mixed, hybrid state $\hat\rho$ in the form
	\begin{align}
		\hat\rho=\int d^2\alpha\sum_{m,n=0}^{d-1} P_{m,n}(\alpha) |\alpha\rangle\langle\alpha|\otimes|m\rangle\langle n|.
	\end{align}
	Note that $P_{m,n}(\alpha)$ can be a complex-valued function for $m\neq n$.
	Moreover, the function fulfills the properties of normalization, $\mathrm{tr}(\hat\rho)=\int d^2\alpha \sum_{n} P_{n,n}(\alpha)=1$, and symmetry, $\hat\rho=\hat\rho^\dag$ $\Leftrightarrow$ $P_{m,n}(\alpha)=P_{n,m}^\ast(\alpha)$, which are necessary for the proper representation of the physical state $\hat\rho$.
	Now, the $P$ distribution conditioned onto the DV state can be written in the form
	\begin{align}\label{eq:neq_DefPMatrix}
		P(\alpha|\hat\Pi)=\mathcal N \vec \psi^\dag\boldsymbol P(\alpha)\vec \psi,
	\end{align}
	where $\boldsymbol P(\alpha)=(P_{m,n}(\alpha))_{m,n}$ and the projection state vector $\vec\psi=(\psi_n)_n$.
	As the normalization constant $\mathcal N$ is positive, we get from Eq. \eqref{eq:neq_DefPMatrix} that $P(\alpha|\hat\Pi)$ is a classical (non-negative) probability distribution for any projection iff the $P$ matrix is non-negative, $\boldsymbol P(\alpha)\geqq 0$ for all $\alpha$.

	As the $P$ function can be highly singular \cite{sperling16}, the matrix $\boldsymbol P(\alpha)$ can be irregular as well.
	For a single CV mode, a nonclassicality-preserving regularization process has been proposed which consists of a convolution of the original $P$~function with a suitable, non-Gaussian kernel $\tilde\Omega(\alpha)$ \cite{kiesel10}.
	For our purposes, this approach can be generalized, yielding the NQP matrix $\boldsymbol P_\Omega(\alpha)=\left[P_{\Omega; m,n}(\alpha)\right]_{m,n}$, with the regular matrix elements
	\begin{align}
		P_{\Omega;m,n}(\alpha)=&\int d^2\alpha'\,\tilde\Omega_w(\alpha-\alpha')P_{m,n}(\alpha'),
	\end{align}
	where our choice of a kernel $\tilde\Omega_w(\alpha)$ is the Fourier transformation of the autocorrelation function
	$\Omega_{w}(\gamma) = \mathcal N_w\int d^2\gamma' e^{-(|\gamma'|/w)^4}e^{-(|\gamma+\gamma'|/w)^4}$
	with a normalization constant $\mathcal N_w$, such that $\Omega_{w}(0)=1$, and $w>0$ denoting the filter width \cite{kiesel10,kiesel11,kiesel11a}.
	In the limit $w\to\infty$, we recover the original $\boldsymbol P(\alpha)$.

	Now, the CHN can be identified with the following necessary and sufficient condition:
	{\it The state $\hat\rho$ shows CHN iff there exists $w>0$ and $\alpha\in\mathbb C$ such that the NQP matrix $\boldsymbol P_\Omega(\alpha)$ is not positive semidefinite,}
	\begin{align}\label{eq:new_NonNegativity}
		\boldsymbol P_\Omega(\alpha)\ngeqq0.
	\end{align}
	We will also use the equivalence of condition \eqref{eq:new_NonNegativity} to the existence of a negative eigenvalue of $\boldsymbol P_\Omega(\alpha)$.

	In another context and restricting to a $2\times 2$ matrix and a convolution with Gaussian kernel yields a Wigner-matrix approach \cite{wallentowitz97}, which inspired the criterion presented here.
	However, the $2\times 2$ Wigner-matrix method has two limitations which we overcome.
	It is restricted to two-level atoms or DV qubit systems.
	More importantly, the Wigner function cannot resolve all nonclassical features \cite{supplement}.

	To experimentally apply condition \eqref{eq:new_NonNegativity}, we reconstruct the NQP matrix with so-called pattern functions \cite{leonhardt95,richter96a}.
	Our data are recorded using balanced homodyne detection, which has been used for state and detector tomography \cite{vogel_book,lvovsky09,grandi16} and the tomography of atomic and optomechanical systems \cite{gunawardena08,gross11,verhagen12}.
	The reconstruction of a DV density matrix in the Fock basis via balanced homodyne detection is well known \cite{dariano95,leonhardt96,richter96b}, and their pattern functions are labeled as $F_{m,n}(x',\varphi')$, where $x'$ is the quadrature for the phase $\varphi'$.
	In the CV scenario, pattern functions $f_{\Omega}(\alpha,w;x,\varphi)$ for the regularized quasiprobabilities $P_{\Omega}$ have been introduced and applied \cite{kiesel11,kiesel11a,agudelo15}.

	We can combine the CV and DV approaches in order to sample the elements of the NQP matrix $\boldsymbol P_\Omega(\alpha)$ from the measured quadratures data points $\{(x_j,\varphi_j,x'_j,\varphi'_j)\}_{j=1}^{N}$,
	\begin{align}
		P_{\Omega;m,n}(\alpha)=&\sum_{j=1}^N \varpi_j f_{\Omega}(\alpha,w;x_j,\varphi_j) F_{m,n}(x'_j,\varphi'_j),
		\label{Eq:Pmn}
	\end{align}
	with weights $\varpi_j\geq 0$ and $\sum_j \varpi_j=1$.
	The full treatment of this technique can be found in the Supplemental Material \cite{supplement} together with a derivation of the weights and the sampling error $\sigma\left[P_{\Omega;m,n}(\alpha)\right]$.
	The introduction of a weighted mean becomes essential for data sets which are not uniformly distributed in phase.
	The weighting corrects for this imperfection, which extends the applicability of our technique beyond previous methods.

\begin{figure}[ht]
	\centering
	\includegraphics[width=0.9\linewidth]{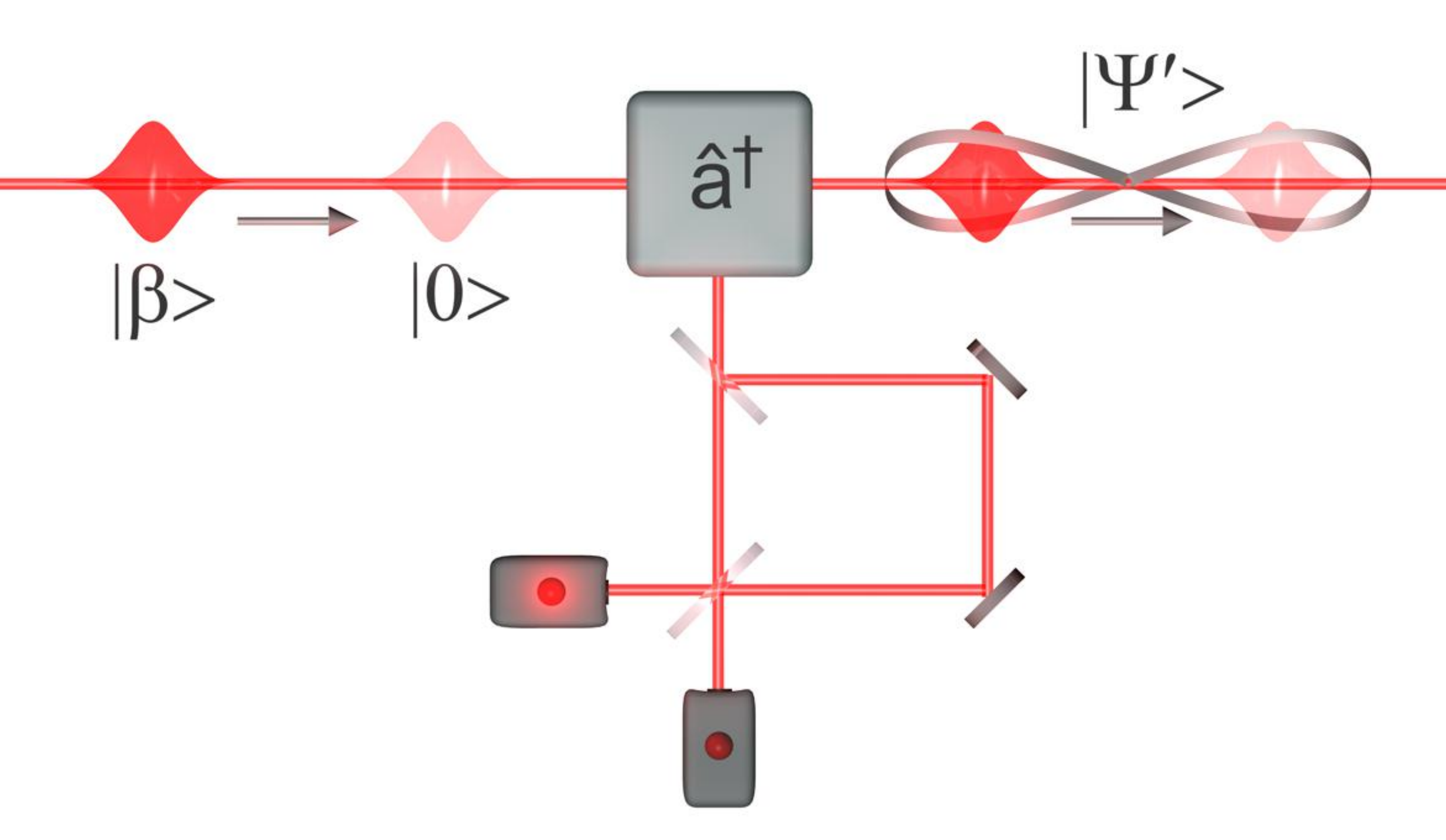}
	\caption{Experimental scheme for the generation of a correlated hybrid state \eqref{eq:target_state}.}
	\label{Fig:ExpScheme}
\end{figure}

\paragraph*{Experimental implementation.---}
	In Fig. \ref{Fig:ExpScheme}, we outline the experiment to produce a state of the type~\eqref{eq:new_CatState}.
	A detailed analysis of the setup may be found in Ref. \cite{jeong14}.
	In our experiment, a single photon-addition device (labeled as $\hat a^\dag$) is fed with two distinct temporal modes containing a coherent and a vacuum state, $|\beta\rangle\otimes|0\rangle$.
	The device realizes a stimulated parametric down-conversion process heralded by the detection of a idler photon; see Ref. \cite{sperling13a} for theoretical details.
	A properly unbalanced Mach-Zehnder interferometer is placed in the idler path to restore the indistinguishability with the temporal modes of the herald photons.

	A click of one of the detectors after the interferometer certifies the addition of a photon in either of the modes, $t\hat a^\dag\otimes \hat 1+r\hat 1\otimes \hat a^\dag$.
	This superposition of creation operations is parametrized with $t$ ($r=\sqrt{1-|t|^2}$), which can be controlled via the relative transmission between the two interferometer arms.
	Choosing $t=1/\sqrt{{|\beta|^2}+2}$ results in correlated signal pulses of the form
	\begin{align}
		|\Psi'\rangle &= 2^{-1/2}\left(|\beta\rangle\otimes|1\rangle+\frac{\hat a^\dagger|\beta\rangle}
		{\sqrt{\langle\beta|\hat a\hat a^\dag|\beta\rangle}}\otimes|0\rangle\right)\nonumber\\
		& \approx 2^{-1/2} (|\beta\rangle\otimes|1\rangle+|g\beta\rangle\otimes|0\rangle).
		\label{eq:target_state}
	\end{align}
	Here the approximation $\hat a^\dag|\beta\rangle\approx|g\beta\rangle$ is used, where the optimal amplitude gain (maximizing the fidelity) is $g=(1+\sqrt{1+4/|\beta|^2})/2$.

	The symmetric target state \eqref{eq:new_CatState} can be simply obtained with a phase-space displacement \cite{jeong14}.
	Since we focus on CHN, which is not affected by such operation, the experimental state \eqref{eq:target_state} with $\beta \cong 1.4$ is analyzed instead.
	Also note that compared to the state \eqref{eq:target_state}, imperfect detectors additionally lead to small DV contributions with photon numbers above one \cite{sperling13a}.

%%%%%%%%%%%%%%%%%%%%%%%%%%%%%%%%%%%%%%%%%%%%%%%%%%%%%%%%%%%%
%ON PURPOSE: FALSE POSITIONING ON PURPOSE -> SEE OUTPUT PDF%
%%%%%%%%%%%%%%%%%%%%%%%%%%%%%%%%%%%%%%%%%%%%%%%%%%%%%%%%%%%%
%xxxxxxxxxxxxxxxxxxxxxxxxxxxxxxxxxxxxxxxxxxxxxxxxxxxxxxxxxxxxxxxxxxxxxxxxxxxxxxx
\begin{figure*}[t]
	\includegraphics[width=0.9\textwidth]{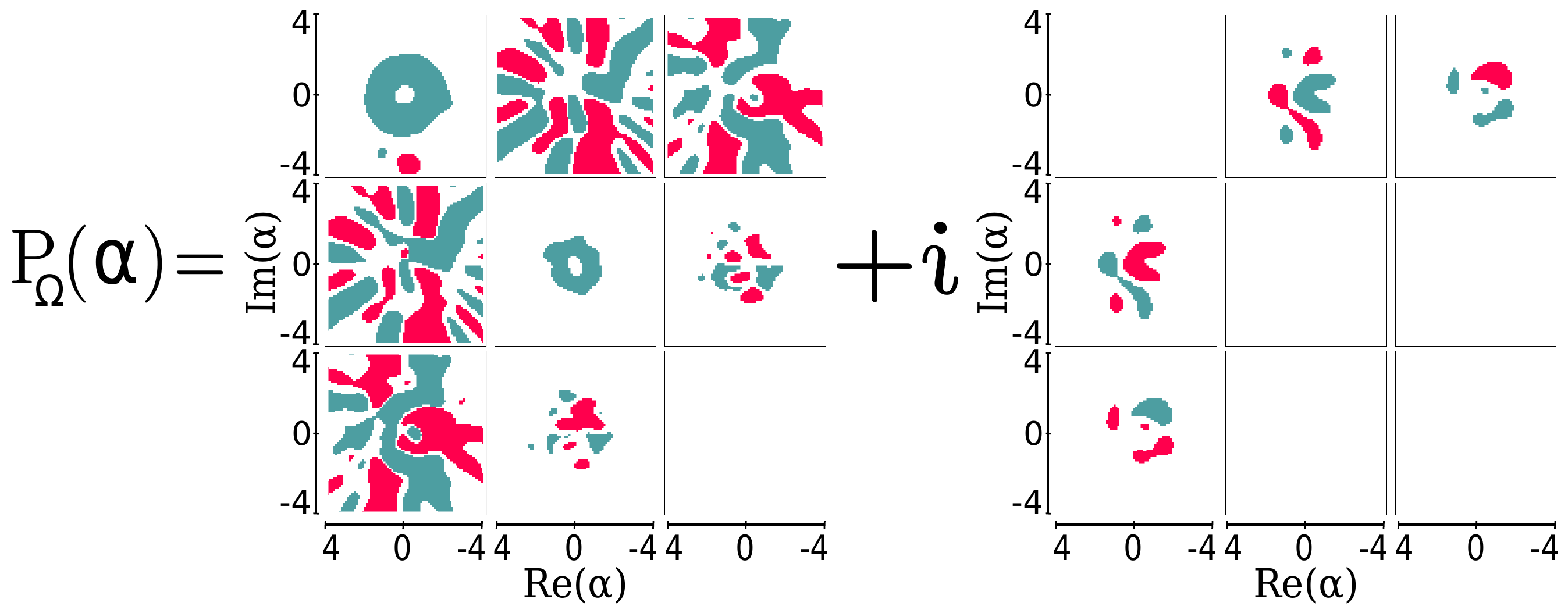}
	\caption{Reconstructed $3\times 3$ NQP matrix $\boldsymbol P_{\Omega}(\alpha)=\left[P_{\Omega;m,n}(\alpha)\right]_{m,n=0,1,2}$.
		Significant values, $|S|>5$, are displayed.
		Green denotes positive values and red negative ones.
		}
	\label{Fig:MatrixP}
\end{figure*}
%xxxxxxxxxxxxxxxxxxxxxxxxxxxxxxxxxxxxxxxxxxxxxxxxxxxxxxxxxxxxxxxxxxxxxxxxxxxxxxx

\paragraph*{Results.---}
	The balanced homodyne detection of the generated state yields an ensemble $\{(x_{j},\varphi_{j},x_j',\varphi_{j}')\}_{j=1}^N$ of $N=372\,000$ data points.
	Based on Eq. \eqref{Eq:Pmn}, we reconstructed the NQP matrix $\boldsymbol P_\Omega(\alpha)$ shown in Fig. \ref{Fig:MatrixP}.
	For the DV part, we present the first three elements, $m,n\in\{0,1,2\}$.
	Higher contributions are not relevant as they have a reconstruction error $\sigma\left[P_{\Omega;m,n}(\alpha)\right]$ that exceeds $34\%$.
	The first observation from Fig. \ref{Fig:MatrixP} is that the reconstruction approximates to some extent our theoretical expectations of the state \eqref{eq:new_CatState}.
	With our sensitive approach, however, we can also identify some deviations, and we can test for CHN in terms of condition \eqref{eq:new_NonNegativity}.
	Let us discuss some results of our analysis.

	In contrast to the ideal state \eqref{eq:new_CatState}, our produced state includes nonzero matrix contributions for more than one photon in the DV mode, e.g., $P_{\Omega;2,2}(\alpha)\not\approx0$.
	Imperfect detectors employed in the addition process can be one source of this behavior \cite{sperling13a}.
	The diagonal elements  also exhibit negative contributions, e.g., $P_{\Omega;0,0}(\alpha)<0$ for $\alpha\approx-4i$, although the projection onto the DV vacuum state is expected to correspond to a classical coherent state.
	More rigorously, the $|n\rangle\langle n|$-conditioned, regularized $P$ functions show a maximal significance of negativities of $S_{0}=8$, $S_{1}=4$, and $S_{2}=3$ standard deviations, where $S_n=\max_\alpha\{-P_{\Omega;n,n}(\alpha)/\sigma(P_{\Omega;n,n}(\alpha))\}$.
	The reason for the significant negativities of $P_{\Omega;0,0}(\alpha)$ is that our technique is sensitive enough to go beyond the approximation $\hat a^\dag|\beta\rangle\approx|g\beta\rangle$ in Eq.~(\ref{eq:target_state}).
	Note, additional analyses \cite{supplement} show that the Wigner function cannot significantly verify this nonclassicality.

	So far we discussed the Fock-diagonal projections of the NQP matrix.
	Now, we also apply our CHN criteria \eqref{eq:new_NonNegativity} for general projections $\vec \psi$ in Eq. \eqref{eq:neq_DefPMatrix}.
	For this purpose, we adopt the eigenvalue approach from Ref. \cite{sperling15}, and we define the submatrices $\boldsymbol P_{\Omega;n}(\alpha)=\left[P_{\Omega;m,m'}(\alpha)\right]_{m,m'=0,\ldots,n}$.
	Hence, we have that $\boldsymbol P_{\Omega;0}(\alpha)$ corresponds to the previously considered $P_{\Omega;0,0}(\alpha)$,
	$\boldsymbol P_{\Omega;1}(\alpha)$ corresponds to the DV subspace in which the state \eqref{eq:new_CatState} lives (i.e., the span of the Fock states $|0\rangle,|1\rangle$),
	and $\boldsymbol P_{\Omega;2}(\alpha)$ is the full matrix shown in Fig.~\ref{Fig:MatrixP}.
	The eigenvector $\vec\psi_{n,\alpha}$ to the minimal eigenvalue of $\boldsymbol P_{\Omega;n}(\alpha)$ describes the optimal projection $\hat\Pi=|\psi\rangle\langle\psi|$ that can be done in the $n$-photon subspace.
	This means, a negative eigenvalue $e_{n,\alpha}=\vec\psi_{n,\alpha}^\dag \boldsymbol P_{\Omega;n}(\alpha)\vec \psi_{n,\alpha}$ certifies the maximal CHN for this point $\alpha$ in phase space.

	The reconstructed NPQ matrix yields the maximal significances of negativities $\Sigma_{n}=\max_\alpha[-e_{n,\alpha}/\sigma(e_{n,\alpha})]$, similarly to $S_n$ for diagonal projections.
	As expected for $n=0$, we observe a CHN with $S_0=\Sigma_0=8$ standard deviations.
	For $n=1$, the off-diagonal contribution $P_{\Omega;0,1}(\alpha)$ has a quite strong impact, which can be seen from the maximal verification of quantumness with $\Sigma_1=32$ standard deviations.
	For comparison, the two possible diagonal projections in this subspace yield only $S_0=8$ and $S_1=4$.
	Hence, the major source of CHN comes from the CV-DV interference terms $P_{\Omega;0,1}(\alpha)$, in which the information on the nonclassical correlations is encoded and which describes the coherent superpositions term $|\beta\rangle\langle{-}\beta|\otimes|0\rangle\langle 1|$.
	The negativity of the full $3\times3$ matrix in Fig.~\ref{Fig:MatrixP} only adds a small contribution to the nonclassicality, i.e., $\Sigma_{{2}}=33\approx\Sigma_{{1}}$.

\paragraph*{Conclusions.---}
	Motivated by the need of CV-DV quantum communication, we introduced the concept of conditional hybrid nonclassicality.
	In contrast to the standard approach of joint correlations, conditional hybrid nonclassicality describes the ability of a CV-DV state to produce a nonclassical state when performing a projecting measurement (i.e., heralding) in one subsystem.
	Beyond the conceptual formulation, we also derived a directly accessible and robust technique to verify the quantumness under study in terms of a regular phase-space matrix.
	The latter functional matrix highlights the interplay between the CV and DV degrees of freedom.
	Negativities in our quasiprobability matrix certify the conditional hybrid nonclassicality.

	We directly implemented our approach by realizing a Schr\"odinger-cat-like state through correlating two temporally separated pulses of light with the help of an interferometric photon-addition process.
	We sampled the nonclassicality quasiprobability matrix and performed a detailed analysis of the state.
	We verified conditional hybrid nonclassicality with high significance.

	The notion of conditional hybrid nonclassicality is a promising candidate for characterizing the usefulness of states for applications at the interface between CV and DV quantum systems.
	It provides a link between the phase-space nonclassicality in quantum optics with the qudit treatment in quantum information science.
	The nonclassicality quasiprobability matrix yields an intuitive understanding of the quantum nature of states and correlations.
	This technique is demonstrated to be experimentally applicable even to imperfect measurements.

\paragraph*{Acknowledgement.---}
	The authors gratefully acknowledge fruitful discussions with B. K\"uhn.
	This work was supported by the Deutsche Forschungsgemeinschaft through SFB 652, Project No. B12.
	J. S. and W. V. acknowledge funding from the European Union's Horizon 2020 research and innovation program under Grant Agreement No. 665148.
	L. S. C., M. B., and A. Z. acknowledge support from Ente Cassa di Risparmio di Firenze and from	the Italian Ministry of Education, University and Research (MIUR), under the ``Progetto Premiale: QSecGroundSpace''.

%\appendix

\section{SUPPLEMENTAL MATERIAL}
	This supplemental material is a guide towards the proper reconstruction the nonclassicality quasiprobability (NQP) matrix elements. 
	It provides some additional features of the conditional hybrid nonclassicality (CHN).
	In a first part, we review some known methods and describe the modifications for the state reconstruction used here.
	In the second part, we derive a treatment for non-uniform phase distributions based on a weighted sampling approach.
	In the third part, an extended discussion of our concepts is presented.
	
\section{Pattern functions}
\label{part:one}
	Consider the physical quantity $F$ and the pair of variables $(x,\varphi)$, quadratures $x$ and phases $\varphi$, that follow the quadrature probability distribution $p(x;\varphi)$, where $\int_{-\infty}^\infty dx\, p(x;\varphi)=1$ for all $\varphi$.
	The $x$ and $\varphi$ values are measured in the range $-\infty<x<\infty$ and $0\le \varphi<\pi$, respectively.
	Then $F$ can be written as
	\begin{equation}
		F =  \int_{-\infty}^{\infty}dx \int_0^\pi d\varphi\, \frac{p(x,\varphi)}{\pi}\, f(x,\varphi),
	\end{equation}
	which means that the quantity $F$ can be determined through the function $f(x,\varphi)$.
	This family of such so-called pattern functions allows us to directly estimate $\overline F$ (i.e., sample the quantity $F$) together with its standard error of the mean $\sigma(F)$ from a given experimental data set $\{(x_j,\varphi_j)\}_{j=1}^{N}$ via
	\begin{align}\label{Eq:estF}
		\overline F=\frac{1}{N}\sum_{j=1}^N f(x_j,\varphi_j),
	\end{align}
	and $\sigma(F)^2=(\overline{F^2}-\overline F^2)/N$ for an equally weighted sampling.
	Note that the denominator $N$ in $\sigma(F)^2$ is typically replaced by $N-1$, the so-called Bessel's correction, which becomes irrelevant for large $N$.
	In this section, the phases are assumed to be uniformly distributed.
	This standard sampling approach will be generalized in the second part of this supplement.

\begin{figure}[ht]
	\includegraphics[scale=0.4]{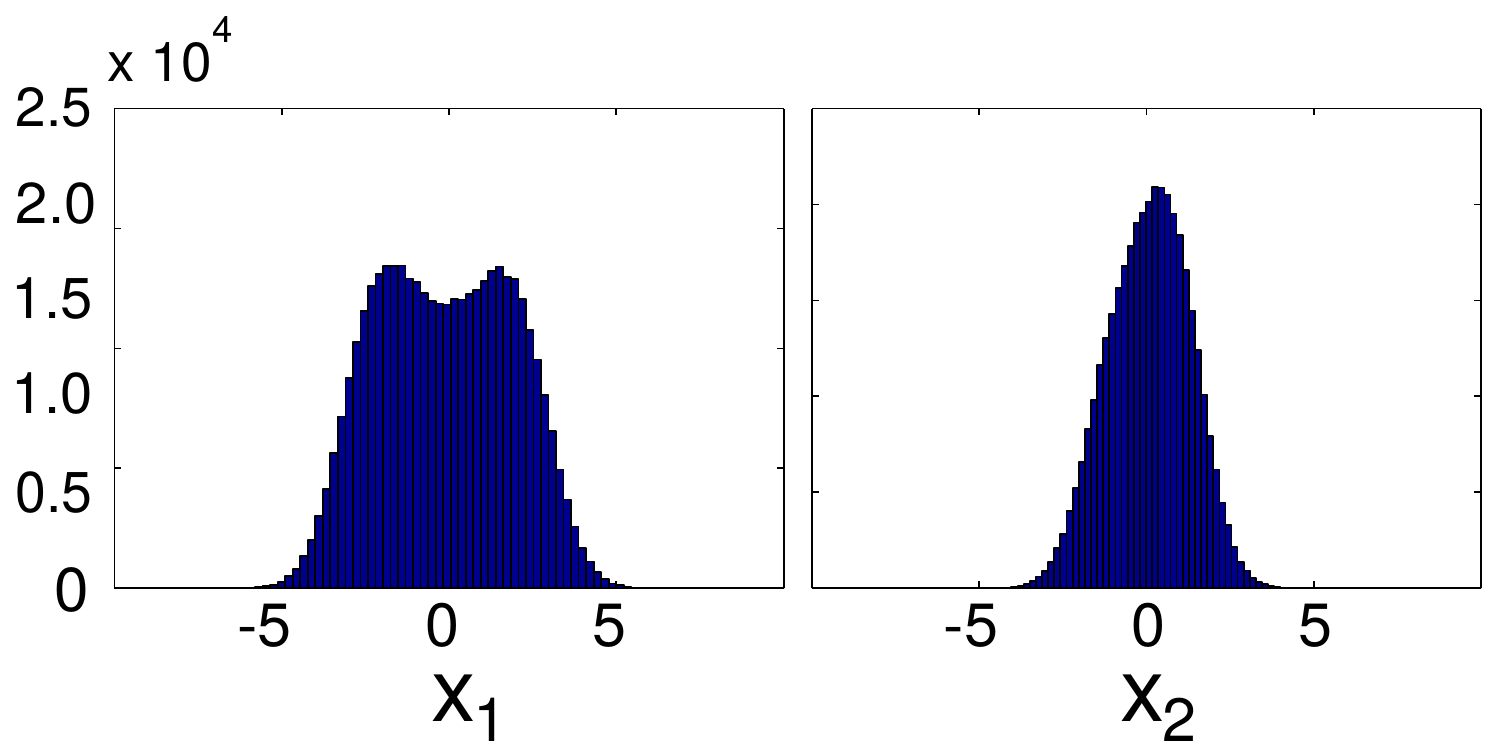}
	\caption{
		The measured counts of the marginal quadrature distributions for the DV mode, $x_1$, and the CV mode, $x_2$.
		The total number of data points is $N=372\,000$.
	}\label{Fig:QDist}
	\end{figure}

	We deal with physical quantities that can be estimated from balanced homodyne detection (BHD).
	This yields the quadrature variances $x=x(\varphi)$, where $\varphi$ is an experimentally adjustable phase-difference between the signal field and the local oscillator.
	The set of data which is obtained from two BHDs, considering a two-mode system, consist in $N$ pairs $\{(x_{1,j},\varphi_{1,j},x_{2,j},\varphi_{2,j})\}_{j=1}^{N}$.
	In our current scenario, the data set is an ensemble of $N=372\,000$ measured quadrature values for an equally spaced set of six fixed phase values per mode.
	A histogram of the measured marginal quadrature distributions for the discrete-variable (DV) mode, $x_1$, and the continuous-variable (CV) mode, $x_2$, is show in Fig. \ref{Fig:QDist}.

\subsection{CV pattern functions}
	
	The pattern functions for a single-mode filtered NQP $P_\Omega(\alpha)$ have been derived in Ref. \cite{kiesel11a}.
	It was shown that
	\begin{equation}
		P_{\Omega}(\alpha) = \int_{-\infty}^{\infty} dx\int_0^\pi d\varphi\,\frac{p(x;\varphi)}{\pi} f_\Omega(x,\varphi;\alpha,w),
	\end{equation}
	with the pattern function
	\begin{align}
		&f_\Omega(x,\varphi;\alpha,w)=
		\\\nonumber
		&\frac{1}{\pi}\int_{-\infty}^{\infty}db \,|b|\, e^{2i b |\alpha| \sin(\arg(\alpha)+\varphi-\pi/2)}e^{ibx}e^{b/2}\Omega_w(b).
	\end{align}
	Here, we restrict ourselves to the filter function $\Omega_{w}(\gamma) =\mathcal N_w\int d^2\gamma' e^{-(|\gamma'|/w)^4}e^{-(|\gamma+\gamma'|/w)^4}$, with $\mathcal N_w$ being chosen such that $\Omega_{w}(0)=1$ and $w>0$ denoting the filter width.
	The nonclassicality quasiprobability, $P_{\Omega}(\alpha)$, is represented as the expectation value of the function $f_\Omega(x,\varphi;\alpha,w)$.
	Analogously to Eq. \eqref{Eq:estF}, when the pairs $(x,\varphi)$ are experimentally measured the expectation value of $P_{\Omega}$ can be replaced by the empirical estimate,
	\begin{equation}
		\overline{P_{\Omega}(\alpha)}= \frac{1}{N}\sum_{j=1}^{N} f_\Omega(x_j,\varphi_j;\alpha,w).
	\end{equation}
	As a set of data for BHD includes in general a large number of data points, several hundred of thousands, the fast evaluation of pattern functions is quite relevant.
	For this purpose, a Fourier technique is applied \cite{kiesel11a}.
	
\subsection{DV pattern functions}
	The reconstruction of the density matrix elements in the Fock representation, $\rho_{m,n}=\langle m|\hat\rho|n\rangle$, from the quadrature component distribution requires another set of pattern functions, $F_{m,n}(x,\varphi)$, which gives
	\begin{equation}\label{Eq:rhomn}
		\rho_{m,n}= \int_{-\infty}^\infty dx\int_0^\pi d\varphi\,  \frac{p(x;\varphi)}{\pi}F_{m,n}(x,\varphi).
	\end{equation}
	The approach to compute the pattern function and their explicit form can be found in Refs. \cite{richter96a,richter99}.
	Decomposing $F_{k,l}(x,\varphi)=f_{k,l}(x)\, e^{i(k-l)\varphi}$, they are given by
	\begin{equation}
		f_{k,l}(x) =i^{k-l} \sqrt{\frac{k!}{l!}} \int_{-\infty}^{\infty}\,du\, |u|\, e^{-u^2/2} u^{l-k}\, L_k^{l-k}(u^2) e^{iux},
	\end{equation}
	where $L^{l-k}_k$ denote the associated Laguerre polynomials.
	Analogously to the CV case, the expectation value can be replaced by the empirical estimate
	\begin{equation}
		\overline{\rho_{m,n}}= \frac{1}{N} \sum_{j=1}^N F_{m,n}(x_j,\varphi_j).
	\end{equation}
	
\subsection{Higher-order, bipartite significances}
	Summarizing the CV and DV sampling approaches, we find that hybrid systems can be described through the elements of the NQP matrix, $P_{\Omega;m,n}(\alpha)$.
	These elements can be sampled according to
	\begin{align}
		\overline{P_{\Omega;m,n}(\alpha)}=\frac{1}{N}\sum_{j=1}^N f_\Omega(x_{1,j},\varphi_{1,j};\alpha,w)F_{m,n}(x_{2,j},\varphi_{2,j}).
	\end{align}
	For classical states, the NQP matrix is non-negative. 
	We can write for a general, Hermitian matrix $\boldsymbol P$ that $\vec v^\dag\boldsymbol P\vec v=e\geq0$, where $\vec v$ is the normalized eigenvector to the minimal eigenvalue $e$ of $\boldsymbol P$.
	Suppose we sample the matrix $\boldsymbol P=\overline{\boldsymbol P}\pm\sigma(\boldsymbol P)$.
	Then we can compute the eigenvector $\vec v$ to the minimal eigenvalue $\overline e$ of $\overline{\boldsymbol P}$---with $\overline e=\vec v^\dag\overline{\boldsymbol P}\vec v$---and we also get the linearly propagated error from $\sigma(e)=|\vec v|^\mathrm{T}\sigma(\boldsymbol P)|\vec v|$ with $|\vec v|=(|v_1|,|v_2|,\dots)^T$ \cite{sperling13a}.

	In addition, let us also consider $\boldsymbol P_n$.
	That is the $n$th principal leading submatrix of $\boldsymbol P$.
	Consequently, we can estimate the minimal eigenvalues, $e_n=\overline{e_n}\pm\sigma(e_n)$.
	In order to provide statistical significance of the reconstructed matrix, we define the higher-order significances of the minimal eigenvalues of NQP matrix as follows:
	\begin{equation}
		\Sigma_n=\frac{\overline{e_n}}{\sigma(e_n)}.
	\end{equation}
	As the minimal bound for the eigenvalues of classical states is $e_\rm{cl} = 0$,
	the absolute value of the significance corresponds to the distance of $e$ to $e_\rm{cl.}$ in units of the error $\sigma(e)$, i.e., $|\Sigma_n| = |e - e_\rm{cl}|/\sigma(e)$.
	The sign of $\Sigma_n$ shows if we are consistent with this bound, $\Sigma_n\geq 0$, or clearly violate it.

\subsection{Pattern functions for a discrete set of phases}\label{subsec:DiscretePhases}
	Ideally, the phases at which the quadratures are measured should be scanned in the whole interval $0\leq \varphi<\pi$ in a uniform distribution.
	In our actual experiment, the quadratures are obtain just at a given number $I$ of fixed, equidistant phases.
	When the sampling function varies rapidly with respect to the phase as in the CV scenario, one can modify the pattern functions \cite{kiesel11a}.
	This reads
	\begin{align*}
		\overline F& =\int_{-\infty}^\infty dx\int_0^\pi d\varphi\,  \frac{p(x;\varphi)}{\pi}\, f(x,\varphi)\\
			& =\sum_{k=1}^{I} \int_{-\infty}^\infty dx \int_{-\frac{\pi}{2I}}^{\frac{\pi}{2I}} d\varphi\,  \frac{p(x;\varphi_k+\varphi)}{\pi}\, f(x,\varphi_k+\varphi)\\
			& \approx\sum_{k=1}^{I} \int_{-\infty}^\infty dx\, \frac{p(x;\varphi_k)}{\pi}\int_{-\frac{\pi}{2I}}^{\frac{\pi}{2I}} d\varphi\,   f(x,\varphi_k+\varphi)\\
			& =\frac{1}{I}\sum_{k=1}^{I} \int_{-\infty}^\infty dx\, p(x;\varphi_k) f'(x,\varphi_k),
	\end{align*}
	with the modified pattern function
	\begin{equation}
		f'(x,\varphi_k) =\frac{I}{\pi}\int_{-\frac{\pi}{2I}}^{\frac{\pi}{2I}} d\varphi\,   f(x,\varphi_k+\varphi).
	\end{equation}
	
\section{Weighted Sampling}
\label{part:two}
\begin{figure}[ht]
	\includegraphics[scale=0.30]{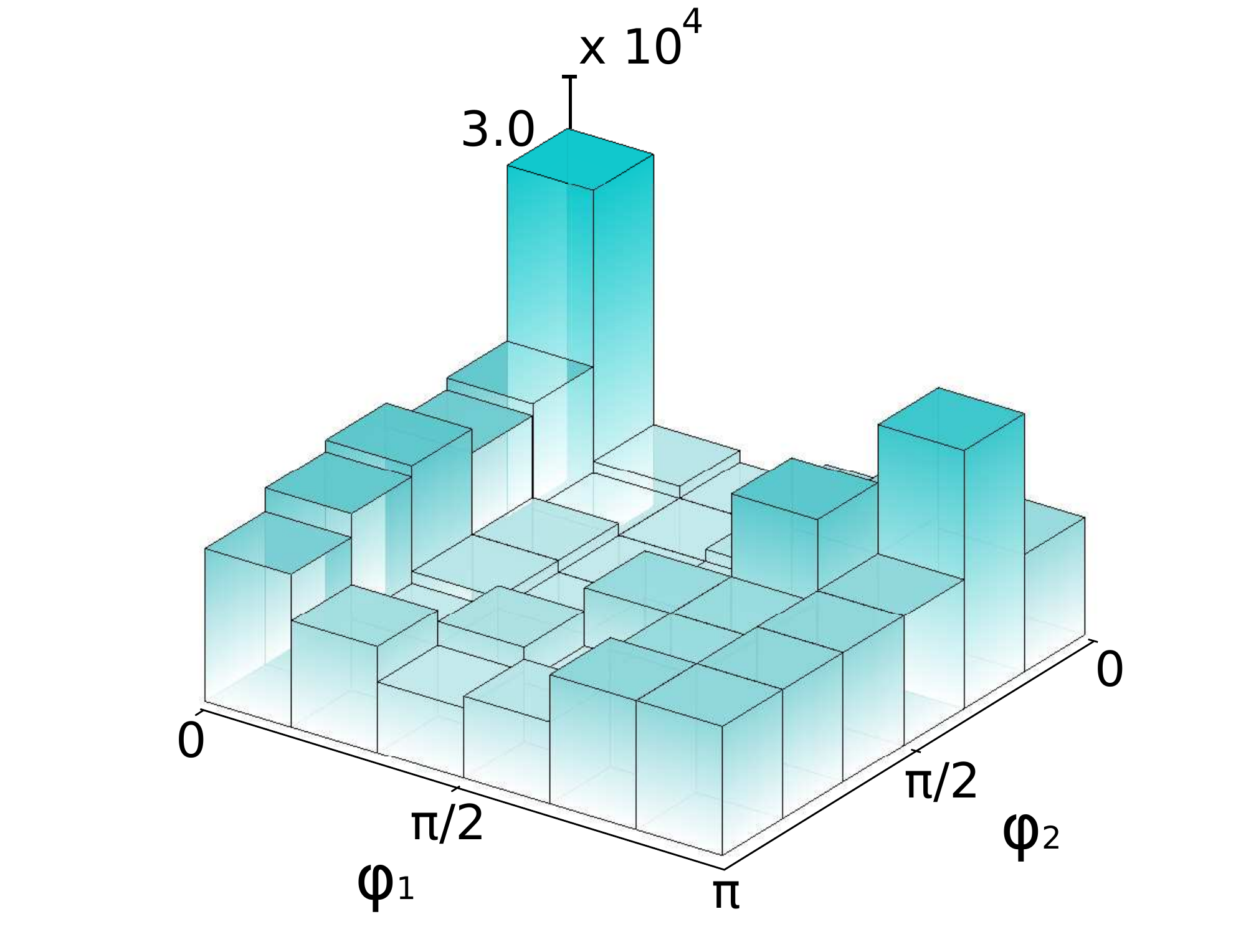}
	\caption{
		The measured counts of the phases for the discrete-variable (DV) mode, $\varphi_1$, and the continuous-variable (CV) mode, $\varphi_2$.
		Our $N=372\,000$ data points are non-uniformly distributed over $I=6\times 6=36$ phase intervals.
	}\label{Fig:PhaseDist}
\end{figure}

	In our scenario, we commonly have sets of data for equidistant, but not uniformly distributed phases; see Fig. \ref{Fig:PhaseDist}. 
	Then the ordinary arithmetic mean for the sampling does not give the correct estimation of the given physical quantities.
	For this reason, we will apply a weighted sampling approach to correct for the unbalanced distribution of data points.
	The general approach of weighted arithmetic means can be found, e.g., in Ref. \cite{bevington69}.
	Let us recall that for a data set $\{(x_j,\varphi_j)\}_{j=1,\ldots,N}$, the considered weights $\varpi_j$ ($\varpi_j\geq0$ and $\sum_{j=1}^N \varpi_j=1$) yield the sampling formula
	\begin{align}\label{eq:weightedM}
		\overline F=\sum_{j=1}^{N} \varpi_j f(x_j,\varphi_j).
	\end{align}
	In particular for $\varpi_j=1/N$, we retrieve the unweighted case in the first part.
	
\subsection{Single-mode case}
	Suppose we have a set of measured data points which is organized in the form $\{(x_j^{(l)},\varphi^{(l)})\}_{j=1,\ldots,N_l;l=1,\ldots,I}$, where $I$ is the number of measured phases and $N_l$ is the number of measured quadratures for the $l$th phase.
	Without a loss of generality, we can assume that the data are ordered with increasing phase, $0\leq\varphi^{(l)}<\varphi^{(l+1)}<\pi$.
	Interpreting a sampling formula in terms of a frequentist probability, we can estimate
	\begin{align}\label{eq:WeightedSampling}
		&F=\int_0^\pi d\varphi\int_{-\infty}^\infty dx\, \frac{p(x;\varphi)}{\pi}f(x,\varphi) \nonumber\\
		\approx&\overline F=\sum_{l=1}^I\sum_{j=1}^{N_l} \varpi_j^{(l)} f(x_j^{(l)},\varphi^{(l)}),
	\end{align}
	where the weightings, $\varpi_j^{(l)}\geq 0$ and $\sum_{l,j} \varpi_j^{(l)}=1$, represent the probability distribution $p(x;\varphi)$ in some limit.
	We define the Heaviside function $\Theta(t)=1$ for $t\geq 0$ and $\Theta(t)=0$ for $t<0$.
	Further note that the probabilities are completely characterized by their cumulative distribution,
	\begin{align}\label{eq:CumulativeDistribution}
		\mathcal P(x\leq X\wedge\varphi\leq \Phi)
		=\int_0^\Phi d\varphi\int_{-\infty}^X dx\, \frac{p(x;\varphi)}{\pi}\nonumber\\
		\approx\sum_{l=1}^I\sum_{j=1}^{N_l} \varpi_j^{(l)} \Theta(X-x_j^{(l)})\Theta(\Phi-\varphi^{(l)}).
	\end{align}

	In the following, we determine the coefficients $\varpi_j^{(l)}$ for the weighted sampling formula \eqref{eq:WeightedSampling} from the cumulative distribution $\mathcal P$ in Eq. \eqref{eq:CumulativeDistribution}.
	We need to satisfy the following two requirements:
	(i) The quadrature density $p(x;\varphi^{(l)})$ for a given phase is approximated by the relative frequencies of measurement outcomes $x_j^{(l)}$ for this phase $\varphi^{(l)}$;
	(ii) The recovered quadrature distribution of phases is uniformly distributed.

	For (i), we consider the conditional probability
	\begin{align}
		\mathcal P(x\leq X|\varphi=\varphi^{(l)})=\frac{\mathcal P(x\leq X\wedge\varphi=\varphi^{(l)})}{\mathcal P(\varphi=\varphi^{(l)})}.
	\end{align}
	We get from our estimation on the one hand
	\begin{align*}
		&\mathcal P(x\leq X|\varphi=\varphi^{(l)})
		=\frac{\int_{-\infty}^X dx\,\frac{p(x;\varphi^{(l)})}{\pi}}{\int_{-\infty}^{\infty} dx\,\frac{p(x;\varphi^{(l)})}{\pi}}
		\\
		=&\int_{-\infty}^X dx\,p(x;\varphi^{(l)})
		\approx\frac{1}{N_l}\sum_{j=1}^{N_l}\Theta(X-x_{j}^{(l)}),
	\end{align*}
	i.e., the normalized sum of all data points $x_j^{(l)}$ below $X$,
	and on the other hand
	\begin{align*}
		&\mathcal P(x\leq X|\varphi=\varphi^{(l)})
		\approx\frac{\sum_{j=1}^{N_l}\varpi_{j}^{(l)}\Theta(X-x_{j}^{(l)})}{\sum_{j=1}^{N_l}\varpi_{j}^{(l)}}.
	\end{align*}
	This means that $\sum_{j}\Theta(X-x_{j}^{(l)})$ is proportional to $\sum_{j}\varpi_{j}^{(l)}\Theta(X-x_{j}^{(l)})$.
	As this relation hold for all $X$, we get that $\varpi_{j}^{(l)}$ has to be constant with respect to $j$,
	\begin{align}
		\varpi_{j}^{(l)}=\varpi^{(l)}.
	\end{align}

	Addressing requirement (ii), we consider a marginal phase distribution $\mathcal P(\varphi^{(l)}\leq \varphi <\varphi^{(l+1)})$ for phases in an interval, which is described by
	\begin{align*}
		&\mathcal P(\varphi^{(l)}\leq \varphi <\varphi^{(l+1)})
		\\
		=&\int_{\varphi^{(l)}}^{\varphi^{(l+1)}} d\varphi\int_{-\infty}^\infty dx\,\frac{p(x;\varphi)}{\pi}
		=\frac{\varphi^{(l+1)}-\varphi^{(l)}}{\pi}
	\end{align*}
	or estimated via
	\begin{align*}
		\mathcal P(\varphi^{(l)}\leq \varphi <\varphi^{(l-1)})
		\approx\sum_{j=1}^{N_l} \varpi^{(l)}=\varpi^{(l)}N_l.
	\end{align*}
	Hence, we conclude
	\begin{align}
		\varpi^{(l)}=\frac{\varphi^{(l+1)}-\varphi^{(l)}}{N_l\pi}.
	\end{align}

	In our case, the phases are equidistant.
	That is, the interval from $0$ to $\pi$ is split into $I$ equally sized intervals which results in $\varphi^{(l+1)}-\varphi^{(l)}=\pi/I$.
	Thus, we find for our measurements that the weighting coefficients for the sampling formula \eqref{eq:WeightedSampling} are
	\begin{align}\label{eq:final_weights}
		\varpi_j^{(l)}=\frac{1}{N_lI}.
	\end{align}

\subsection{Bipartite hybrid sampling}
	For our particular case of a bipartite system, we get the following sampling formula of discrete phases that are not uniformly distributed:
	\begin{equation}\label{eq:allSample1}
		\overline{P_{\Omega;m,n}(\alpha)}= \sum_{l=1}^{I}\sum_{j=1}^{N_l}
		\frac{1}{IN_l}g_j^{(l)},
	\end{equation}
	where we use
	\begin{align}
		g_j^{(l)}=f'_\Omega(x_{1,j}^{(l)},\varphi_{1}^{(l)};\alpha,w)\,F'_{m,n}(x_{2,j}^{(l)},\varphi_{2}^{(l)})
	\end{align}
	and $\{(x_{1,j}^{(l)},\varphi_{1,j}^{(l)},x_{2,j}^{(l)},\varphi_{2,j}^{(l)})\}_{j=1,\ldots,N_l;l=1,\ldots,I}$ defines our two-mode data set.
	It is worth mentioning that the total number of data points is $N=\sum_{l=1}^I N_l$.
	Here, instead of each of the data points contributing equally to the final average, we have a weighted mean of the product $g_j^{(l)}$ of pattern functions per phases pair.
	In Fig. \ref{Fig:PhaseDist}, we showed the corresponding distribution of data points in the two-dimensional phase intervals which yields the weightings in Eq. \eqref{eq:final_weights}.
	Let us also stress that the pattern functions are independent of our weighting coefficients, cf. Eq. \eqref{eq:weightedM}.
	Further and as it was similarly shown in the previous subsection, Eq. \eqref{eq:allSample1} represents a proper estimate which satisfies our requirements (i) and (ii).

	In the following, we derive the sampling-error estimation for the expression \eqref{eq:allSample1}.
	The standard approach is the treatment of all $g_j^{(l)}$ as random variables which are independent and distributed according to a normal distributions with a variance $\sigma(g_j^{(l)})^2$.
	This gives 
	\begin{align*}
		\sigma(P_{\Omega;m,n}(\alpha))^2=\sum_{l=1}^I\sum_{j=1}^{N_l} \left(\frac{1}{I N_l}\right)^2\sigma_j^{(l)2}.
	\end{align*}
	For a fixed phase $\varphi^{(l)}$, the random variables $g_j^{(l)}$ are identically distributed ($\sigma_j^{(l)}=\sigma^{(l)}$) which allows us to rewrite
	\begin{align}\label{eq:weightedsamplingerror}
		\sigma(P_{\Omega;m,n}(\alpha))^2=\frac{1}{I^2}\sum_{l=1}^I \frac{\sigma^{(l)2}}{N_l},
	\end{align}
	where the empirical variance for a fixed phase is the standard estimate $\sigma^{(l)2}=\sum_{j=1}^{N_l} g_j^{(l)2}/N_l-\big(\sum_{j=1}^{N_l} g_j^{(l)}/N_l\big)^2$.
	Equation \eqref{eq:weightedsamplingerror} is the sampling error for the expression in Eq. \eqref{eq:allSample1} for a non-uniform distribution of phases.

%==========================================================================================================================================================
\section{Conditional nonclassicality}
\label{part:three}

%%%%FALSE POSITIONING ON PURPOSE
\begin{figure*}
	\center
	\includegraphics[scale=0.6]{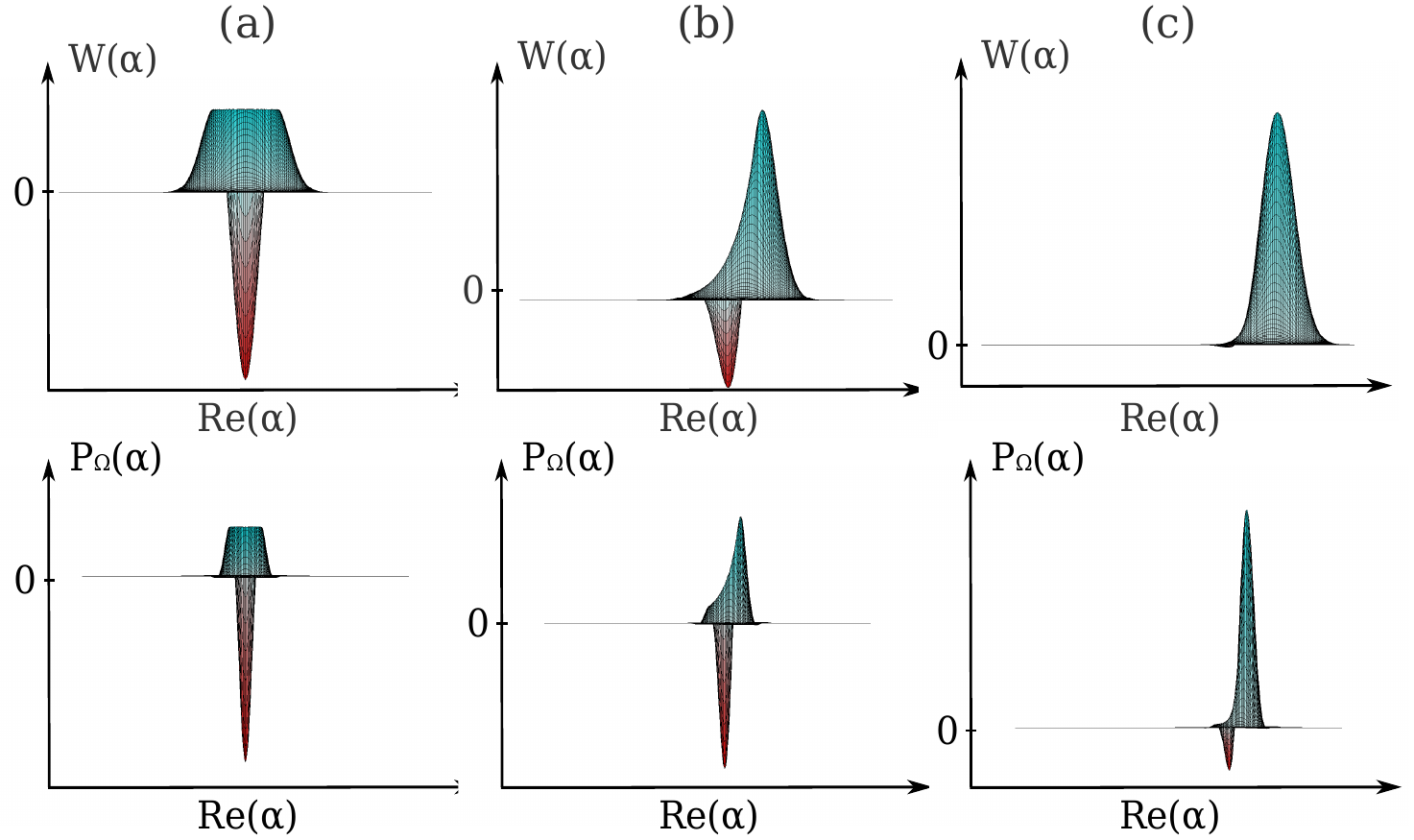}
	\caption{
		The Wigner function $W(\alpha)$ (top row) and the regularized $P$ function $P_\Omega(\alpha)$ (width $w=1.9$, bottom row) are shown for the state $\hat a^\dagger|\beta\rangle$ and different $\beta$ values: column (a) $\beta=0$, column (b) $\beta=0.9$, and column (c) $\beta=2.6$.
	}
	\label{Fig:SPACS}
\end{figure*}

\subsection{Interpretation of the CHN notion}
	The CHN provides a link between DV and CV systems.
	Its definition states: The conditional Glauber-Sudarshan $P$ distribution of the CV subsystem has to have no counterpart in classical statistics for a projection $\hat \Pi$ in the DV subsystem, i.e., $P(\alpha|\hat \Pi)\ngeqq 0$.
	This means that the hybrid state can be used to generate a nonclassical CV state through a measurement in the DV system.
	This can be naturally interpreted in terms of the heralded generation of nonclassical light.

	An important remark is that any arbitrary entangled state will be represented by a NQP matrix which is not positive-semidefinite.
	This can be seen from the fact that any state which does not exhibit CHN takes the form
	\begin{align}\label{eq:nonCHN}
		\hat\rho=\int dP_\mathrm{cl}(\alpha,\phi) |\alpha\rangle\langle\alpha|\otimes|\phi\rangle\langle\phi|,
	\end{align}
	where $|\alpha\rangle$ is a CV coherent state, $|\phi\rangle$ is an arbitrary DV state, and $P_\mathrm{cl}$ is a classical probability distribution (i.e., $P_\mathrm{cl}\geqq0$) over the pure product states $|\alpha\rangle\otimes|\phi\rangle$.
	Note, states of the form \eqref{eq:nonCHN} describe the convex hull over all pure states without CHN, $|\alpha\rangle\otimes|\phi\rangle$.
	Therefore, such states without CHN are automatically separable.
	Conversely, an entangled state cannot take the form \eqref{eq:nonCHN}.

	To prove that CHN can exist beyond entanglement, let us consider initially a two-mode squeezed-vacuum state (likewise, EPR state)
	\begin{align}
		\hat\rho=\sum_{n,m=0}^\infty (1-p)\sqrt{p^{n+m}} \ket{n}\ket{n}\bra{m}\bra{m},
		\label{Eq:TMSV}
	\end{align}
	with $0<p<1$ and where $p=\tanh^2\xi$ is related to the squeezing parameter $\xi$ of the initial input fields, where we choose the phase such that $\xi>0$.
	Note that this requires a generalization of the dimensionality of the DV system to $d=\infty$.
	The state \eqref{Eq:TMSV} is entangled and, therefore, exhibits CHN.
	For instance, if we perform a conditional measurement onto the $n$th photon-number state, $\hat \Pi=|n\rangle\langle n|$, we get the conditional state $\hat\rho|_{\hat\Pi}=|n\rangle\langle n|$.
	For $n>0$, this describes a nonclassical $n$-photon state of light, proving CHN.

	Performing a dephasing operation on the state \eqref{Eq:TMSV}, we observe a decay of entanglement \cite{sperling12}.
	In particular, a full dephasing results in a phase-randomized two-mode squeezed-vacuum state,
	\begin{align}
		\hat\rho'=\sum_{n=0}^\infty (1-p)p^n |n\rangle\langle n|\otimes|n\rangle\langle n|,
		\label{Eq:FRTMSV}
	\end{align}
	whose nonclassical features have been extensively studied in Ref. \cite{agudelo13}.
	For instance, this non-Gaussian state has a non-negative Wigner function, and the full phase diffusion made the initial EPR state separable.
	Still, when heralding onto the $n$th Fock state, we also get $\hat\rho'|_{\hat\Pi}=|n\rangle\langle n|$.
	Again, this state is nonclassical for $n>0$, and we confirmed CHN.
	Hence, CHN can exist without entanglement---or, entanglement is not needed to herald nonclassical states, such as single photons.
	
	Also note that there is a difference between the definition of the conditional $P$ function [Eq. (5) in the Letter], further defining CHN, and the derived criterion to probe CHN [Eq. (7) in the Letter].
	The conditional $P$ function is defined in terms for each projective measurement $\hat\Pi$.
	But the NQP matrix includes the information about all projections in the DV mode.

\subsection{Regularized \textit{P} function vs. Wigner function}

	To further support the claim that our approach gives more insights into the nonclassicality compared to previous approaches, see for example Ref. \cite{wallentowitz97}, we consider the conditional state $\hat\rho_{|0\rangle\langle 0|}$ as an example; cf. Eq. (3) in the Letter.
	Further on, we have also outlined that the generated hybrid state is approximated by $\hat\rho=|\Psi'\rangle\langle\Psi'|$, with $|\Psi'\rangle \approx 2^{-1/2} (|\beta\rangle\otimes|1\rangle+|g\beta\rangle\otimes|0\rangle)$; cf. Eq. (9) in the Letter.
	In particular, we used the approximation $\mathcal{N}^{1/2}\,a^\dagger|\beta\rangle\approx |g\beta\rangle$, with a proper normalization constant $\mathcal N$.
	Thus, the actual conditional state reads
	\begin{align}\label{eq:SPACS}
		\hat\rho|_{|0\rangle\langle 0|}=\mathcal N \hat a^\dag|\beta\rangle\langle\beta|\hat a.
	\end{align}
	In Fig. \ref{Fig:SPACS}, we compare the Wigner and the regularized $P$ function of the conditional state \eqref{eq:SPACS} for different displacements $\beta$.
	Note, this phase-space function corresponds to the element $P_{\Omega;0,0}(\alpha)$ of our regularized NQP matrix, which we reconstructed from our data [Fig. 2 in the Letter].

	From Fig. \ref{Fig:SPACS}, we observe for both phase-space quasi\-probabilities, $W(\alpha)$ and $P_\Omega(\alpha)$, that the height of the negative part relative to the positive part decreases with increasing coherent amplitudes $|\beta|$.
	For larger $|\beta|$ values, the Wigner function is unable to resolve those negativities in the presence of unavoidable reconstruction errors.
	In the same scenario, the regularized $P$ function has still a significant negative part.

	The used approximation for $\mathcal N^{1/2}\hat a^\dag|\beta\rangle$ is aimed at representing---to some extend---the coherent state $|g\beta\rangle$.
	Specifically, this would imply a non-negative Wigner function, which was also experimentally verified in Ref. \cite{jeong14}.
	However, the photon-added coherent state $\hat a^\dag|\beta\rangle$ is only an approximation to a coherent state which is revealed in the form of negativities of the regularized $P$ function---for our data, with a statistical significance of $S_{0}=8$ standard deviations under the same experimental conditions as in Ref. \cite{jeong14}.
	Therefore, the applied technique of regularized $P$ function is a highly-sensitive technique beyond the standard Wigner approach.


\begin{thebibliography}{999}
\bibitem{comment1}
	One could also choose $d=\infty$, e.g., $|\psi\rangle=\sum_{m=0}^\infty \psi_m|m\rangle$ in the discrete photon-number basis.
	This requires the additional restriction to normalized vectors, $\langle \psi|\psi\rangle=\sum_{m=0}^\infty|\psi_m|^2=1$, in contrast to unnormalized eigenvectors of CV observables, e.g., the momentum operator.
\bibitem{comment2}
	The initial example of a hybrid state $|\Psi_\mathrm{cat}\rangle\otimes|q\rangle$ consists of one CV mode and a four-dimensional DV mode which consists of two qubits.
	The projection operator is defined via the two-qubit state $|\psi\rangle=(|0\rangle\otimes|0\rangle+|1\rangle\otimes|1\rangle)/\sqrt 2$.
\end{thebibliography}
\end{document}